\documentstyle[aps,epsf,graphicx]{revtex}
\newcommand{\beq}{\begin{equation}}
\newcommand{\eeq}{\end{equation}}        
\newcommand{\bqa}{\begin{eqnarray}}        
\newcommand{\eqa}{\end{eqnarray}}

\newcommand{\UKcol}{{\sf UKQCD collaboration}}

\newcommand{\Expt}[1]{\langle #1 \rangle}
\newcommand{\heading}[1]{\noindent{\bfseries #1}\newline}
\newcommand{\Sw}{{\cal S}}
\newcommand{\Ha}{{\cal H}}

\newcommand{\Op}{{\cal O}}
\newcommand{\Ave}[1]{\overline{\cal #1}}
\newcommand{\W}{{\cal W}_{\Box}}
\newcommand{\DU}{\left[dU\right]}

\newcommand{\Dpf}{\left[ d\phi \right]}
\newcommand{\Dpfd}{\left[ d\phi^{\dagger} \right]}

\newcommand{\Dpi}{\left[d\pi\right]}
\newcommand{\Var}[1]{\sigma^2 \left( #1 \right) }
\begin{document}

\draft

\title{ 
\hfill\begin{minipage}{0pt}\scriptsize \begin{tabbing}
\hspace*{\fill} Edinburgh preprint 98/13\\ 
\hspace*{\fill} LTH  428\\
\end{tabbing} 
\end{minipage}\\[8pt]  
Parallel Tempering in Lattice QCD 
with O(a)--Improved Wilson Fermions 
}

\author{\frenchspacing B\'alint Jo\'o, Brian Pendleton, Stephen M. Pickles and  Zbigniew Sroczynski} 
\address{Department of Physics and Astronomy,
	The University of Edinburgh, \\
        The King's Buildings, Edinburgh EH9 3JZ,
        Scotland}
        
\author{\frenchspacing Alan C. Irving}
\address{Theoretical Physics Division,
         Department of Mathematical Sciences\\
         University of Liverpool,
         PO Box 147, Liverpool L69 3BX, UK}

\author{\frenchspacing James C. Sexton} 
\address{School of Mathematics,
         Trinity College and Hitachi Dublin Laboratory,
         Dublin 2,
         Ireland}

\author{\UKcol{}}
\maketitle

\begin{abstract}
We present details of our investigations of the Parallel Tempering
algorithm. We consider the application of action matching technology
to the selection of parameters. We then present a simple model of the
autocorrelations for a particular parallel tempered system.  Finally
we present results from applying the algorithm to lattice QCD with
O(a)--improved dynamical Wilson Fermions for twin sub--ensemble systems.

\pacs{12.38.Gc, 11.15.Ha, 02.70.Lq}

\end{abstract}

\section{Introduction}
\label{s:Introduction}
The computational cost of lattice QCD has always been 
enormous. During the last few years the power of supercomputers
has grown immensely but simulations with dynamical fermions
are still very time consuming. 

One of the most popular algorithms for dynamical fermion simulations
is the Hybrid Monte Carlo (HMC) algorithm \cite{BJPHmc}. However it
has been suggested that HMC is not very efficient at decorrelating
some long range observables such as the topological
charge~\cite{Boyd1}. On the other hand, results from the SESAM
collaboration~\cite{SESAM} indicate that HMC simulations using Wilson
fermions seem to tunnel between topological sectors at an adequate
rate. The results of SESAM indicate an autocorrelation time for the 
topological charge of about 50 HMC trajectories.

With such high computational costs it is always necessary to keep 
an eye open for alternative algorithms. Parallel Tempering (PT) 
or Exchange Monte Carlo was proposed in ~\cite{Hukushima} 
to assist decorrelation in spin--glass systems. A lucid description of 
PT and related algorithms such as Simulated Tempering and their
applications to spin--glass and other systems may be found  in \cite{Marinari,MariPari}. 

Recently PT has been applied to simulations of lattice QCD with staggered fermions \cite{Boyd2} and this preliminary study indicated that the autocorrelation 
times for some observables were significantly improved over the
normal HMC results.

In this paper we present our study of the PT algorithm using 2
flavours of degenerate O(a)--improved Wilson fermions \cite{Clover}
with a non-perturbatively determined coefficient \cite{Jansen}.

PT simulates several lattice QCD ensembles concurrently, hereafter
referred to as {\em sub--ensembles}, with different
simulation parameters. PT exploits the fact
that the equilibrium distributions of the configurations in individual
sub--ensembles have an overlap, and occasionally tries to swap
configurations between pairs of sub--ensembles, while keeping all 
sub--ensembles in equilibrium.  This is done in such a way that the 
factorisation of the joint equilibrium distribution of configurations
into the individual distributions for each sub--ensemble is not
disturbed by the swapping.

The acceptance of these swap attempts depends on how close the sub--ensembles
are to each other in parameter space. The concept of distance in parameter
space is formalised in \cite{ACIJCS,ACIlat97,Match} by 
the machinery of action and observable matching. In theory, this 
technology should allow the selection of an optimal set of parameters
to maximise the swap acceptance rate between the sub--ensembles.

Another possibility is to use the action matching technology to define
curves in parameter space on which some observable such as
$r_0$~\cite{R0} is constant. PT can, in principle be used to
simulate numerous points on such a curve in one simulation. However it
must be stressed that this scenario is different from the one above.
Matching observables is not the same as matching the
action~\cite{Match}. Hence in this case one does not in general have
as good control over the swap acceptance rate as in the situation outlined
previously.

The remainder of this paper is organised as follows.

Our particular variant of the PT algorithm is described in detail
in the following section, where we show that it satisfies
detailed balance and present a formula for the acceptance rate
of the swap attempts. We then relate this formula to 
the distance in parameter space as defined in the context of action
matching technology.

The swapping of configurations between sub--ensembles is expected to 
reduce the autocorrelation times of observables within individual 
sub--ensembles with respect to their HMC autocorrelation times. In section
\ref{s:Autocorrelations} we discuss the simple case of a PT system 
consisting of two sub--ensembles. We suggest a model for the
autocorrelation function in the PT sub--ensembles in terms of that of
the  original HMC ensembles.
 
Our simulations are discussed in Section \ref{s:SimulationDetails} and
our results are presented in Section \ref{s:Results}. We show that
indeed our acceptance rate formula of section \ref{s:Algorithm} is
borne out by the simulation results. We estimate the autocorrelation
time of the plaquette for several swap acceptance rates and compare
these estimates with the prediction of the model outlined in section
\ref{s:Autocorrelations}.

Our summary and conclusions are presented in Section \ref{s:Conclusions}. 

\section{The Parallel Tempering Algorithm}
\label{s:Algorithm}

\medskip
\heading{Notation}
Let each sub--ensemble be labelled by an index $i$ and let 
the phase space of sub--ensemble $i$ be $\Gamma_i$.
Each sub--ensemble has an action 
$\Sw_i$ which depends upon the set of parameters and the fields
of the sub--ensemble.

We simulate two flavours of dynamical fermions using the 
standard pseudofermionic action:
\beq
\Sw_i = -\beta_i\W(U) +  \phi^{\dagger} \left( M^{\dagger}(\kappa_i, c_i)
M(\kappa_i, c_i) \right)^{-1} \phi \label{e:Action}
\eeq
where $\W$ is the Wilson plaquette action, $U$ are the gauge fields,
$\phi$ are the pseudofermion fields, and $M$ is the O(a)--improved
fermion matrix with hopping parameter $\kappa$ and clover coefficient
$c$. In addition, for HMC algorithms we need to introduce momentum
fields $\pi_i$ and construct Hamiltonian functions $\Ha_i = \pi^2_i +
\Sw_i$. A state in sub--ensemble $i$ is then represented by the triple
$a_i=\left(U_{i}, \pi_{i}, \phi_{i} \right)$ while the parameter set
for sub--ensemble $i$ is the triple of real numbers $\left( \beta_i,
\kappa_i, c_i \right)$. Note that the subscript $i$ serves only to distinguish ensembles and will be dropped when discussing a single sub--ensemble.

Each sub--ensemble has the phase space
\beq
\Gamma_i = \{U_i\} \otimes \{\pi_i\} \otimes \{\phi_i\}.
\eeq
We note at this stage that all the $\Gamma_i$ are identical and
the only distinguishing features of individual ensembles are their
parameter sets and quantities which depend upon these such
as $\Sw_i$ or $\Ha_i$.

A PT simulation state is thus the collection of states $\{ a_i | i=1...n\}$,
where $n$ is the number of sub--ensembles. The overall PT phase space is
the direct product of the phase spaces of the sub--ensembles
\beq
\Gamma_{\rm PT} = \prod_{i=1}^{n} \Gamma_i.
\eeq

\medskip 
\heading{Detailed Balance}
In a PT simulation one needs to construct a Markov process which 
has (joint) equilibrium probability distribution:
\begin{equation}
P^{\rm eq}_{\rm PT} = \prod_i P^{\rm eq}_i(U, \pi, \phi) \label{e:conv_criterion}
\end{equation}
where $P^{\rm eq}(U,\pi,\phi)$ is the desired equilibrium probability distribution of 
the individual sub--ensemble $i$. In our case 
\begin{eqnarray} 
P^{\rm eq}_i(U,\pi,\phi) &=& {1\over Z_i} e^{-\Ha_i(U,\pi,\phi)} \\           
       Z_i  &=& \int \DU \Dpi \Dpf \Dpfd e^{-\Ha_i(U,\pi,\phi)}.
\end{eqnarray}

Equation \ref{e:conv_criterion} formalises our notion of simulating
ensembles independently. To be more precise, the Markov steps
within any individual sub--ensemble are independent of those in the
others, but the resulting sub--ensembles
are not independent as they are coupled by the swapping steps.
However
the overall joint equilibrium distribution of the PT system is 
not affected by the swapping, and remains the product of the individual
equilibrium distributions of the sub--ensembles.

We define two kinds of Markov transitions:

\begin{enumerate}
\item {\bfseries Transitions within a single ensemble:}
These transitions can be made with any desired Markovian update
procedure that satisfies detailed balance with respect to $P^{\rm eq}$
for its sub--ensemble. In our case such transitions are made with 
HMC. We refer to the set of HMC trajectories that
are performed between swaps as an HMC step.
\item {\bfseries Transitions between sub--ensembles:}
These transitions are used to connect the phase spaces of 
the sub--ensembles. Such a transition 
would be a proposed swap between any two sub--ensembles $i$ and $j$.
Let $a$ be a configuration in sub--ensemble $i$ and $b$ 
be a configuration in sub--ensemble $j$.
The swap transition can be denoted:
\beq
(a,b) \rightarrow \left\{ 
\begin{array}{l l}
(b,a) & \mbox{if swap is accepted} \\
(a,b)& \mbox{if swap is rejected}. 
\end{array} \right.
\eeq
Let us denote by $P_s(i,j)$ the probability that the swap succeeds.
The detailed balance condition is:
\beq
P_s(i,j)e^{-\Ha_i(a)}e^{-\Ha_j(b)} 
= P_s(j,i)e^{-\Ha_j(a)}e^{-\Ha_i(b)}
\eeq
as the contributions from the other ensembles cancel on both sides.
A suitable choice for $P_s$ is the simple Metropolis \cite{Metropolis}
acceptance probability:
\beq
P_s(i,j) = \mbox{min}\left( 1, e^{-\Delta \Ha} \right) \label{e:AccRate}
\eeq
where 
\beq
\Delta \Ha = \left\{ \Ha_j(a) + \Ha_i(b) \right\} 
            -\left\{ \Ha_i(a) + \Ha_j(b) \right\}
\eeq
which satisfies the detailed balance condition by construction.
\end{enumerate}

The required overall Markov transition should be constructed
of a number of both kinds of transitions. HMC steps within all the 
sub--ensembles are necessary and sufficient for convergence. Transitions between sub--ensembles are not essential but without them PT would basically be the same as running several independent HMC simulations.

\medskip
\heading{Swap Acceptance Rate}
Any extra decorrelation of observables in PT over and above
normal HMC must necessarily come from the swapping transitions.
Control of the acceptance rate for swapping transitions is therefore
important. The swapping probability is determined by the energy
change $\Delta \Ha$ as in (\ref{e:AccRate}). The acceptance
rate for Metropolis-like algorithms of this kind is easily shown 
to be~\cite{Karsch}
\begin{equation}
\langle A \rangle = \mbox{erfc}\left( {1\over2}\sqrt{ \langle \Delta \Ha \rangle} \right). \label{e:AcceptRate}
\end{equation}
Here $\Expt{\Delta \Ha}$ is the average of $\Delta \Ha$ over all 
swap attempts, and $\Expt{A}$ is the average acceptance rate of 
the swap attempts.

\medskip
\heading{Action Matching} 
The action matching formalism outlined in \cite{ACIJCS} formalises
the meaning of distance in parameter space. We review here the salient
points of the discussion.

Let $S_{1}[U]$ and $S_{2}[U]$ be the actions of two lattice gauge theories
with the same gauge configuration space, so that the partition function 
of each is:
\beq 
        {\cal Z}_i=\int \DU \exp \{ -S_{i}[U]\}
\eeq
and the expectation of an observable $\Op$ in ensemble $i$ is:
\beq
\Expt{\Op}_i = \frac{1}{{\cal Z}_i} \int \DU \Op(U) \exp\{ -S_{i}[U] \}.
\eeq
Naturally, using actions dependent on other fields will complicate
the integration measure. If one deals with pseudofermions
for example, these would have to be integrated also, both in the 
partition function ${\cal Z}$ and in the expectation value of observables.
We will not explicitly write out integrations over pseudofermions in 
expectation values, except cases where such an ommission may 
lead to ambiguities.

The expectation of $\Op$ in the other ensemble is given to first order in 
a cumulant expansion by:
\beq
\Expt{\Op}_2 = \Expt{\Op}_1 + \Expt{\tilde{\Op} \tilde{\Delta}_{12}}_1 + ...
\eeq
where $\Delta_{12} \equiv S_1 - S_2$ and $\tilde{\Op} \equiv \Op - \Expt{\Op}$ etc.

The distance between the two actions is  defined as the variance
\beq
        d \equiv \Var{ \Delta_{12} } \equiv \Expt{ \tilde{ \Delta}_{12}^2 }
\eeq
where the expectation is to be evaluated in either sub--ensemble.
 
Three matching conditions have been identified:
\begin{itemize}
\item 
Match the values of observables i.e. require that $\Expt{\Op}_1 = \Expt{\Op}_2$
\item
Minimise $d$
\item
Maximise the acceptance in an exact algorithm for $S_2$ constructed
via accept/reject applied to configurations generated with action $S_1$.
\end{itemize} 

It was shown in \cite{ACIJCS} that the last two conditions
are equivalent to lowest order in a cumulant expansion.
Under special circumstances the first condition is also equivalent to the
other two to lowest order. The prescriptions differ in a calculable way at the next order.

We are now ready to make the connection between PT and the
formalism of action matching. We note that the energy 
difference before and after a PT swap attempt is 
\beq
\delta = \Delta \Ha.
\eeq
The momentum fields cancel exactly in the Hamiltonian
terms and one can deal directly with the actions:
\beq
\delta = S_1(U_2,\phi_2) + S_2(U_1,\phi_1) - S_1(U_1, \phi_1) - S_2(U_2,\phi_2)
\eeq
Collecting the terms depending on the same fields one obtains:
\beq
\delta = \Delta_{12}(U_2,\phi_2) - \Delta_{12}(U_1,\phi_1).
\eeq
We now identify $\delta$ with $-\delta$ in (3.15) in \cite{ACIJCS}. Following
the analysis of \cite{ACIJCS} one may obtain the acceptance rate formula
of the action matching mechanism
\beq
\Expt{A} = \mbox{erfc} \left({1 \over 2} \sqrt{\sigma^2(\Delta_{12})} \right)\,.
\eeq
One can then deduce that
\beq
      \sigma^2(\Delta_{12}) = \Expt{\Delta \Ha} \approx {1 \over 2} \sigma^2(\Delta \Ha) \label{e:Connection}
\eeq
where the second approximate equality is required to derive the
acceptance rate (\ref{e:AcceptRate}).


Our PT parameters were tuned using the action 
matching technology to maximise the acceptance between two 
subensembles using the action
\beq
\Sw_i = -\beta_i \W - T_i   \label{e:Tracelog}
\eeq
with 
\beq 
T_i = {\rm Tr}\,\ln ( Q^{-1}_i )
\eeq
and 
\beq
Q_i = \left( M^{\dagger}(\kappa_i)M(\kappa_i) \right)^{-1}.
\eeq
The tuning was carried out before performing the PT simulation
using configurations from a preliminary HMC run at the desired 
reference parameter set.

However our PT simulations were carried out using the action
\beq
\Sw_i = -\beta_i \W + \phi^{\dagger} Q_i \phi.  \label{e:pfaction}
\eeq
We have found that tuning parameters 
using (\ref{e:Tracelog}) for which we had reliable technology did 
not optimise the swap acceptance of our simulations. The reasons 
for this are discussed below.

Consider first the distance $\sigma^2$ between actions $\Sw_i$ where 
the $\Sw_i$ are as given by (\ref{e:Tracelog}). Then
\beq
\Delta_{12} = \Delta \beta \W + \Delta T
\eeq
with
\bqa
\Delta \beta &=& \beta_2 - \beta_1 \\
\Delta T &=& T_2 - T_1. 
\eqa
The variance of $\Delta_{12}$ in an individual subensemble is 
\beq
\sigma^2(\Delta_{12})_i = \Expt{ ( \Delta \beta \widetilde{ \W } +\widetilde{ \Delta T } )^2 }_i. \label{e:TrLnsigma}
\eeq
One can see that for a given $\Delta T$ one can tune $\Delta \beta$ to 
minimise this variance. 

However when one examines the case of the pseudofermionic 
action of (\ref{e:pfaction}) one finds that 
\beq
\Delta_{12} = \Delta \beta \W  + \phi^{\dagger} \left( Q_1 - Q_2 \right) \phi.
\eeq
When calculating 
the variance of $\Delta_{12}$ one encounters the quadratic term
\beq
\Expt{ \phi^{\dagger}(Q_1 - Q_2) \phi \phi^{\dagger}(Q_1 - Q_2) \phi }_i.
\eeq
This term gives rise to both connected and disconnected pieces when
the integration over the pseudofermion fields is carrried out
\beq
\Expt{ \phi^{\dagger}(Q_1 - Q_2) \phi \phi^{\dagger}(Q_1 - Q_2) \phi }_i = \Expt{ {\rm Tr}^2((Q_1 - Q_2)Q^{-1}_{i}) }^U_i + \Expt{ {\rm Tr}(Q_1 - Q_2)Q^{-1}_i(Q_1 - Q_2)Q^{-1}_i }^U_i.
\eeq                                                    
Here the superscript $U$ on the expectations indicates that they are now to be 
carried out over the gauge fields only.
Hence one finds that 
\beq
\sigma^2_i(\Delta_{12}) = \Expt{ ( \Delta \beta \widetilde{\W} + \widetilde{ {\rm Tr}((Q_1 - Q_2)Q^{-1}_i) } )^2 }^U_i + \Expt{ {\rm Tr}(Q_1 - Q_2)Q^{-1}_i(Q_1 - Q_2)Q^{-1}_i }^U_i   \label{e:pfsigma}
\eeq
We also note that to first order in $Q_1 - Q_2$
\beq
\Delta T \approx {\rm Tr}((Q_1 - Q_2)Q^{-1}_i)
\eeq                                           
                              
Comparing equations (\ref{e:TrLnsigma}) and (\ref{e:pfsigma}) it can be seen 
that using a pseudofermionic action gives rise 
to a connected piece in $\sigma^2_i(\Delta_{12})$ which one  
would not get using the action of (\ref{e:Tracelog}). This connected
piece cannot be tuned away by changing $\Delta \beta$ and it increases the distances in parameter space compared
to when the action of (\ref{e:Tracelog}) is used. If parameters are 
tuned using the action of (\ref{e:Tracelog}) and the simulation is  
carried out using pseudofermions the acceptance rate of the 
swaps will not be optimised.

With hindsight it may be said that using pseudofermions was not the 
best choice for performing our simulations, and that the action of 
(\ref{e:Tracelog}) should have been evaluated on our swap attempts 
to calculate $\Delta_{12}$ instead of using the pseudofermionic action
to calculate $\Delta \Ha$. 

\section{Autocorrelations}
\label{s:Autocorrelations}
\heading{The cost of measuring observables}
The gain from PT is expected to come from the swapping of configurations
between sub--ensembles. This reduction in autocorrelation time is supposed 
to occur due to the fact that the sub--ensembles are 
simulated (between swaps) with independent Markov processes. However
the swaps couple the ensembles and include cross correlations between 
them. Thus care must be taken when using results from separate subensembles
together.

According to \cite{Adk,Sokal} if successive measurements of $\Op$ are
correlated, the sample mean $\Ave{O}$ is given (we use the convention
of \cite{Adk}) by the formula:
\beq
\Ave{O} = \Expt{\Op} \pm \sqrt{\frac{2\tau_{\Op}+1}{N}\Var{\Op}}.
\eeq
Here, $\Var{\Op}$ is the variance of operator $\Op$ given by
\beq
\Var{\Op} = \Expt{\Op^2} - \Expt{\Op}^2
\eeq
and $\tau_{\Op}$ is the integrated autocorrelation time, defined as
\beq
\tau_{\Op} = \sum_{t=1}^{\infty} C_{\Op}(t)
\eeq
and where $C_{\Op}(t)$ is the normalised autocorrelation function:
\beq
C_{\Op}(t) = \frac{1}{\Var{\Op}}\Expt{ \left( \Op(t + I)-\Expt{\Op} \right)\left( \Op(I) - \Expt{\Op} \right)}
\eeq
and the expectation values are over all pairs of $\Op_i$ separated
by an interval t. From now on we shall drop the subscript $\Op$ from these
formulae except where necessary. Furthermore the term `autocorrelation
time' will always be used to refer to the integrated autocorrelation time.

The practical meaning of the statements above is that $2\tau + 1$ 
correlated measurements of $\Op$ are needed in order to reduce the 
error in $\Ave{\Op}$ by the same amount as if two uncorrelated measurements
were used. Markov methods in general produce correlated sequences 
of configurations, and hence correlated sequences of measured observables.
The integrated autocorrelation time $\tau$ is therefore an important
indicator of the performance of a Monte Carlo simulation that is carried out
with the intention of measuring observable $\Op$. 

In particular, if one assumes that the autocorrelation function decays
exponentially
\beq
C(t) = \exp\{-kt\} \label{e:EXPAF}
\eeq
with $k>0$, one finds that
\beq
\exp\{-k\} = \frac{\tau}{\tau + 1} \label{e:INVERT}
\eeq
which is a result we shall use later.

\medskip
\heading{Autocorrelations in twin sub--ensemble PT}
We are interested in whether or not PT will reduce the integrated
autocorrelation time of an observable measured on an ensemble with
some parameter set relative to the corresponding autocorrelation time
of the same observable measured on an ensemble generated at the same
parameters using HMC.  We refer to the former of these autocorrelation
times as the PT autocorrelation time and the latter as the HMC
autocorrelation time.

Let us examine the situation of a PT system with two sub--ensembles.
Sub--ensemble 1 has the desired parameter set, and the other
sub--ensemble has its parameters chosen so as to give some acceptance rate
$\Expt{A}$. We assume that the HMC autocorrelation functions of both
ensembles are the same.  We demonstrate in section
\ref{s:Results} that over the distances in parameter space for which
we can use PT, and with the statistics available, we cannot
differentiate between the autocorrelation times of the plaquette operator
between sub--ensembles, so we regard the above assumption as reasonable.
 
Having made the above assumption, the changes in the 
autocorrelation time due to PT are now controlled solely by the 
number of successful swaps between the sub--ensembles. The swap 
probability in general depends on the particular PT state at which the
swap is attempted, but for simplicity we assume that we can replace individual
swap probabilities with the overall average swap probability which is none 
other than the acceptance rate $\Expt{A}$.

Let the HMC autocorrelation function be denoted $C_{H}(t)$, and the 
PT autocorrelation function of the sub--ensemble of interest be denoted
$C_{PT}(t)$. Consider the connected autocorrelation function:
\beq
        C_{H}(t)=\frac{1}{\Expt{\Op^2}}\sum_{i=0}^{n-t} \Op_{i+t}\Op_{i} 
\eeq
where $n$ is the number of samples of $\Op_i$.

The autocorrelation function in the PT ensemble of interest
can now be written as: 
\beq
C_{PT}(t) = \frac{1}{\Expt{\Op^2}} \left\{ S_{e}+S_{o} \right\} \label{e:DEF}
\eeq
where
\bqa
S_{e} = \sum_{\rm even}\Op_{i+t} \Op_{i}  \label{EVEN} \\
S_{o} = \sum_{\rm odd}\Op_{i+t} \Op_{i}  \label{ODD}\, . 
\eqa
By the even sum we mean that the only terms contributing 
to the sum are those where an even number of swaps succeeded out of the
$t$ tried between the measurements of $\Op_{i+t}$ and $\Op_{i}$.

Given some configuration in one sub--ensemble, after an odd number of 
successful swaps it can only be in the other one. As the HMC steps 
are independent in different sub--ensembles, we expect (to a first 
approximation) no correlation between configurations in a sub--ensemble 
that are separated by an odd number of swaps. Hence we assume that $S_{o}$
sums to zero and we consider only the $S_{e}$ term. 

We then rewrite (\ref{e:DEF}) as:
\beq
C_{PT}= P_{e}C_{H}(t)
\eeq
where $P_{e}$ is the probability that an even number of successful
swaps occur in $t$ trials. $P_{e}$ is given by
\beq
P_{e} = \sum_{i} C^{t}_{i}(1-\Expt{A})^{t-i}\Expt{A}^{i} \label{e:PROB}
\eeq
where the index $i$ runs from $0$ to the largest even integer less than 
or equal to $t$, $i$ is even and $C^{t}_{i}$ is the number of ways 
of choosing $i$ swaps from $t$.

Carrying out the sum in equation (\ref{e:PROB}) one finds 
\beq
P_{e} = \frac{1}{2} \left\{ 1 + \left( 1 - 2\Expt{A} \right)^{t} \right\}
\eeq
leading to the result:
\beq
C_{PT}(t) = \frac{1}{2} \left\{ 1 + \left( 1 - 2\Expt{A} \right)^{t} \right\} C_{H}(t)
\, .\label{e:RESULT}
\eeq
We consider three separate cases.
\begin{enumerate}
\item[i)] {$\mathbf \Expt{A}=0$: \ }
In this case $C_{PT}(t)=C_{H}(t)$, which is what we expect when we do not
carry out any successful swaps. 
\item[ii)] {$\mathbf 0 < \Expt{A} \le \frac{1}{2}$: \ \ }
In this case $C_{PT} \in [\frac{1}{2}C_{H},C_{H})$ and we can see a
reduction in the autocorrelation function of at most a factor of 2.
\item[iii)] {$\mathbf \frac{1}{2} < \Expt{A} \le 1$: \ \ }
In this case the term $ \left( 1 - 2\Expt{A} \right)^{t} $ in 
equation (\ref{e:RESULT}) becomes oscillatory.
In particular if $\Expt{A}=1$ (every swap succeeds) it is
impossible to get an even number of successful swaps out of an odd number of
trials, whereas it is a certainty for an even number of trials.
\end{enumerate} 

If one models the autocorrelation function by an exponential decay
as in (\ref{e:EXPAF}), it is possible to calculate the PT integrated autocorrelation time for the ensemble:
\bqa
\tau_{PT} &=& \sum_{1}^{\infty} C_{PT}(t) \\
     &=& \frac{1}{2}\tau_{H} + \frac{1}{2}\sum_{1}^{\infty}\left( (1-2\Expt{A})\exp\{-k\}\right)^{t} \\
     &=& \frac{\tau_H \left[ 1 + \Expt{A}\left(\tau_{H} -1 \right) \right]}{1+2\Expt{A}\tau_H} \label{e:AUTORES}
\eqa
where the last line follows from using (\ref{e:INVERT}), 
summing the resulting geometric series and simplifying.
The ratio of $\tau_{PT}$ to $\tau_{H}$ is then:
\beq
\frac{\tau_{PT}}{\tau_{H}} = \frac{1+\Expt{A}(\tau_H-1)}{1+2\Expt{A}\tau_{H}}. \label{e:RATIO_PRED}
\eeq

We remark on several features of the ratio in (\ref{e:RATIO_PRED}).
\begin{enumerate}
\item[i)]
When $\Expt{A}=0$, one is, in effect, carrying out two uncoupled 
HMC simulations and the autocorrelation times in each sub--ensemble
remain the same as they would be for HMC simulations.
\item[ii)]
For a fixed $\Expt{A} \in (0,{1 \over 2})$ increasing $\tau_H$ from $0$
has the effect that the ratio of (\ref{e:RATIO_PRED}) approaches the value 
of $1 \over 2$ {\em from above}. The closer $\Expt{A}$ is to $1 \over 2$, the faster this limit is approached. If one is interested in both subensembles
this is still a gain. If one of the two ensembles serves only to 
decorrelate the other and is not otherwise interesting (it is thrown away 
at the end) then one would lose over HMC as one would have done twice the 
work, but gained less than a factor of two.
\item[iii)]
For $\Expt{A}={1 \over 2}$ the ratio is exactly $1 \over 2$ and a breakeven is 
reached, in the sense that one does the work of  
two simulations, but in each subensemble the integrated autocorrelation is 
halved. This is the stage when a sub--ensemble which originally
served no other purpose than to help decorrelate the other one may be 
thrown away without losing out.
\item[iv)]
For $\Expt{A} \in ({1\over2},1]$ the ratio approaches $1 \over 2$ rapidly {\em from below}. In 
this case one clearly wins even if one is only interested in a single 
sub--ensemble. However the gain is not much, as for any reasonable value of 
$\tau_H$ the ratio will have already approached the asymptotic limit 
of $1 \over 2$ to a good level of accuracy.  
\end{enumerate}

One can therefore win most with PT when the acceptance rate is very high, 
and the observable of interest has a very short autocorrelation time.
In such a situation it is possible to gain more than a factor 
of two over the HMC autocorrelation time in each ensemble if the swap
acceptance rate is greater than $1 \over 2$. However if an observable has 
such a short HMC autocorrelation time, it may not be worthwhile employing PT.
Parallel tempering was supposed to be used to decorrelate
observables with long autocorrelation times.
In a typical situation, it would be expected that the gain in each ensemble 
is very close to a factor of 2.

\medskip
\section{Simulation Details}
\label{s:SimulationDetails}

Our PT simulations were carried out on the Cray T3E in Edinburgh. Code for performing the HMC trajectories 
was taken from the GHMC code written for the UKQCD Dynamical Fermions
project, described in \cite{StephsAndZbysh}. 

\medskip
\heading{Program Features}
The PT code ran trajectories on each sub--ensemble in {\em series}. 
Sub--ensembles were labelled from $0$ to $N-1$, where $N$ was the 
total number of sub--ensembles. Swaps of configurations between 
sub--ensembles were attempted according to a boolean
{\em plan matrix} $M$. If, after carrying out the HMC trajectories in 
sub--ensemble
$i$, the element $M_{ij}$ was found to contain {\em true}, the 
code would attempt to swap configurations $j$ and $j+1$. ($j \in [ 0, N-2]$) 
The default matrix had all its elements set to {\em false} except for the 
last row which had all its elements set to {\em true}. This way the 
program would perform all the HMC trajectories on all the ensembles and 
would then attempt a chain of pairwise swaps.

The number of HMC trajectories per sub--ensemble was controlled 
through an independent parameter file for each sub--ensemble. 
This way a sub--ensemble could be equilibrated with the GHMC code and if desired, they could easily be  
taken and further simulated separately using the GHMC code.
Likewise each sub--ensemble kept a separate set of log files for the 
plaquette and for solver statistics. The overall driver routine
kept a log file of the success or failure of swap attempts and the swap
energies.

\medskip
\heading{Simulation Parameters}
Five PT simulations $S1$, $S2$, $S3$, $S4$ and $S5$ were performed,
each of which comprised two sub--ensembles.
The parameters for these simulations are shown in table \ref{t:2PARAMS}.
In all five simulations one sub--ensemble had parameters 
$(\beta=5.2, c=2.0171, \kappa=.13300)$. 
The parameters for the second sub--ensemble were given by action matching 
for $S1$, $S2$ and $S3$, while for $S4$ and $S5$ only $\kappa$ was varied. Thus
we could investigate the PT swap acceptance rate for different distances
in parameter space.

We also had data from a previous HMC simulation with parameters $(\beta=5.2, c=2.0171, \kappa=.13300)$ on lattices of volume $8^3 \times 16$ and $8^3 \times 24$. 

The results from the reference run on the $8^3 \times 16$ lattice
were used to validate the PT code. Our PT simulations 
were also carried out on lattices of this size. Furthermore, it was possible
to compare the autocorrelation times of the plaquette from this 
HMC run with the autocorrelation times of the plaquette from the first 
sub--ensembles of the PT runs.
For the second sub--ensembles, the GHMC code was used only to achieve equilibration.
Thus there is insufficient data to calculate the HMC autocorrelation 
times of the second sub--ensembles.

In the PT simulations each HMC step was one trajectory long. The plan 
matrix used was the default one described earlier. Simulations
$S1$, $S2$ and $S3$ ran for 6000 swap attempts giving 6000 
trajectories for each sub--ensemble, while $S4$ and $S5$ ran for 
only 1000 swap attempts due to time constraints.
 
The matching procedure was performed using with the reference HMC results
from $8^3 \times 24$ lattices, using the methods outlined in \cite{Match}.


\medskip
\heading{Analysis}
We examined the acceptance rate as a function of the average swap
energy change $\Expt{\Delta \Ha}$, and of $\Delta \kappa=\kappa_2 -
\kappa_1$, the change in the hopping parameters. We investigated the
autocorrelation time of the average plaquette.

Errors in ensemble averages were estimated using the bootstrap method.
Autocorrelations were estimated using the sliding window scheme of 
{\em Sokal et al.} \cite{Sokal}.   

\section{Results}
\label{s:Results}
A summary of our results is shown in table \ref{t:MainRes}. We show
for each simulation $\Delta \beta = \beta_2 - \beta_1$, the
corresponding $\Delta \kappa$, $\Expt{\Delta \Ha}$, the acceptance
rate $\Expt{A}$, the integrated autocorrelation time $\tau$ for the
plaquette in sub--ensemble 1 and the autocorrelation time in
sub--ensemble 1 divided by the HMC autocorrelation time,
${\tau_1}/{\tau_{\rm HMC}}$.



\begin{figure}
\begin{center}
\epsffile{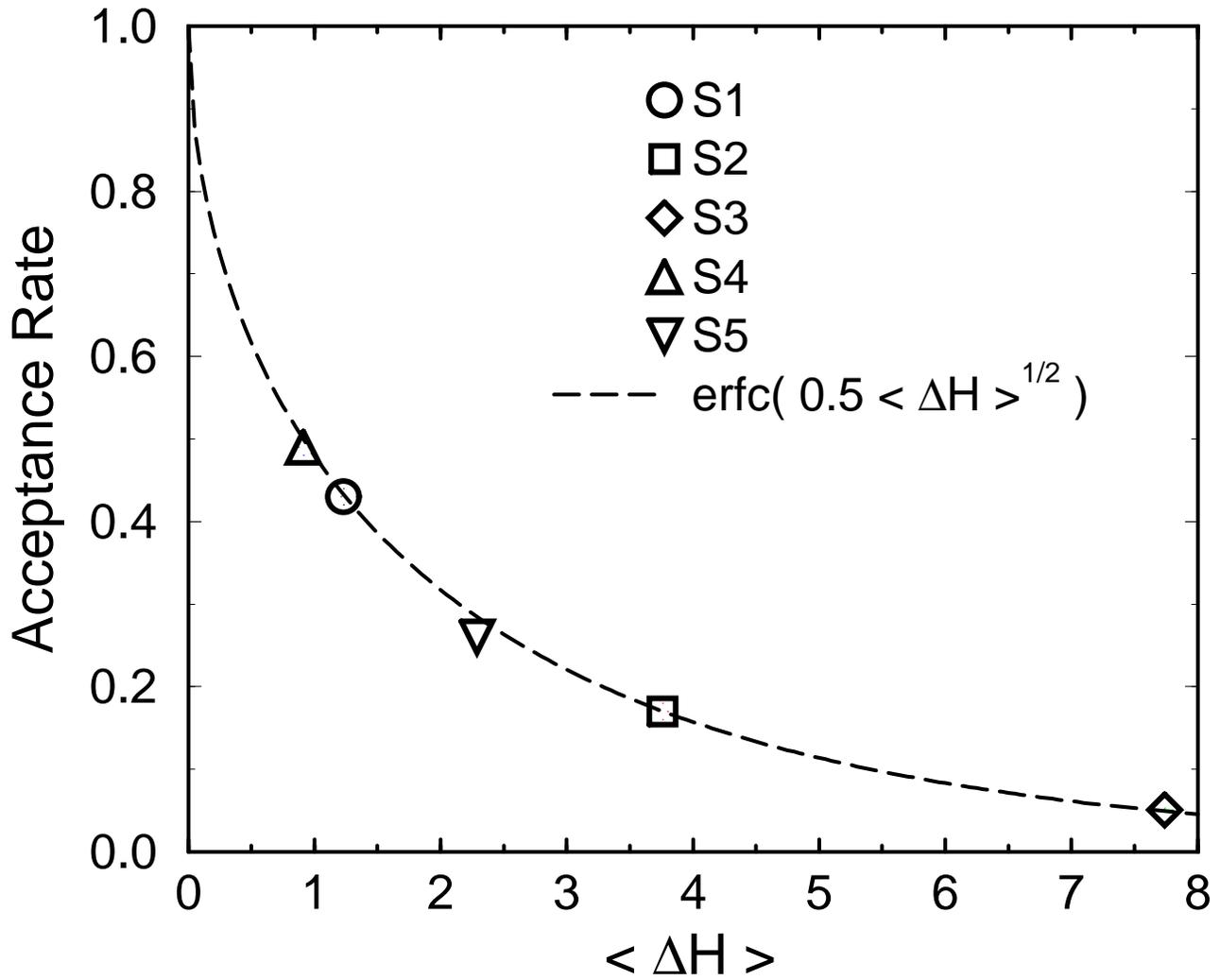}
\caption{Acceptance rate against $\langle \Delta \Ha \rangle$. Error
bars are smaller than the symbols}
\label{f:Acceptance}
\end{center}
\end{figure}

\medskip
\heading{Swap Acceptance Rate} 
Figure \ref{f:Acceptance} shows the measured swap acceptance rates of
the simulations.  The solid line is the acceptance rate formula in
(\ref{e:AcceptRate}). It can be seen that the measured results are in
excellent agreement with its predictions.

\medskip
\heading{Calibration and Matching}
It can be seen from table \ref{t:MainRes} that simulations $S2$ and $S3$ which 
had parameters given by matching the ${\rm Tr}\,\ln$ actions of (\ref{e:Tracelog}) have
lower acceptance rates than $S4$ and $S5$ for which tempering was carried
out only in $\kappa$. We expect that this is due to the noise term
of (\ref{e:pfsigma}) and is the result of using the pseudofermion
action for calculating the swap energy differences. 

To see how large the effect of this noise term is, we can compare the
residual variance $\sigma^2(\Delta_{12})$ from the matching procedure
\cite{Match}, using the ${\rm Tr}\,\ln$ action with the variance as measured in
our PT simulations through $\Expt{\Delta \Ha}$.  Note that we only
have biased estimators for $\sigma^2(\Delta_{12})$ from the matching
procecure, and that we have calculated the residual variance estimate
only for $\Delta \kappa=.0005$.

Table \ref{t:Comppf} contains our comparison of the $\mbox{Tr}\,\ln$ matching
predictions and pseudofermionic measurements for simulation $S3$. We
can see in column 2, our biased estimate of the residual variance on
matching and in column 4 the corresponding predicted acceptance
rate. In column 3 we see the actual variance as measured in the
simulation and in
column 6 the corresponding measured acceptance rate. We expect the
difference in the variances to be due to the four point term in
equation (\ref{e:pfsigma}). We can therefore numerically estimate
the four point term to be 
\begin{equation} 
\Expt{ {\rm Tr}(Q_2 - Q_1)Q^{-1}_i(Q_2 - Q_1)Q^{-1}_i }^U_i = 6.6(2)
\end{equation} 
for simulation $S3$. 

Note that if during our swap acceptance steps, we 
would discard the pseudofermion fields, and calculate the energy
change using the $\mbox{Tr}\,\ln$ action by the methods outlined in \cite{Match},
we would suffer a workload hit, but would expect an accept rate of 
around $48\%$ in the case of simulation $S3$. Thus using pseudofermions
was a poor way to proceed originally. However as the action difference
scales like the lattice volume, going to larger lattices would effectively
cancel all the gain one could obtain by using the $\mbox{Tr}\,\ln$ action to evaluate the
swap action/energy difference.

\medskip
\heading{Autocorrelation Times and Efficiency}
The autocorrelation times of the plaquette operator on the
sub--ensembles with parameter $\kappa=.1330$ are shown in column 5 of
table \ref{t:MainRes}.  We also show for comparison the
autocorrelation time estimated from our independent HMC run at the
same parameter set. In table \ref{t:IndepHMC} we gather some estimates
of the integrated autocorrelation time of the plaquette for some
independent HMC runs at similar parameters to our PT runs.  It can be
seen that the HMC autocorrelation times agree with each other within
estimated errors, justifying the assumptions of our model of section
\ref{s:Algorithm}.

\begin{figure}
\epsffile{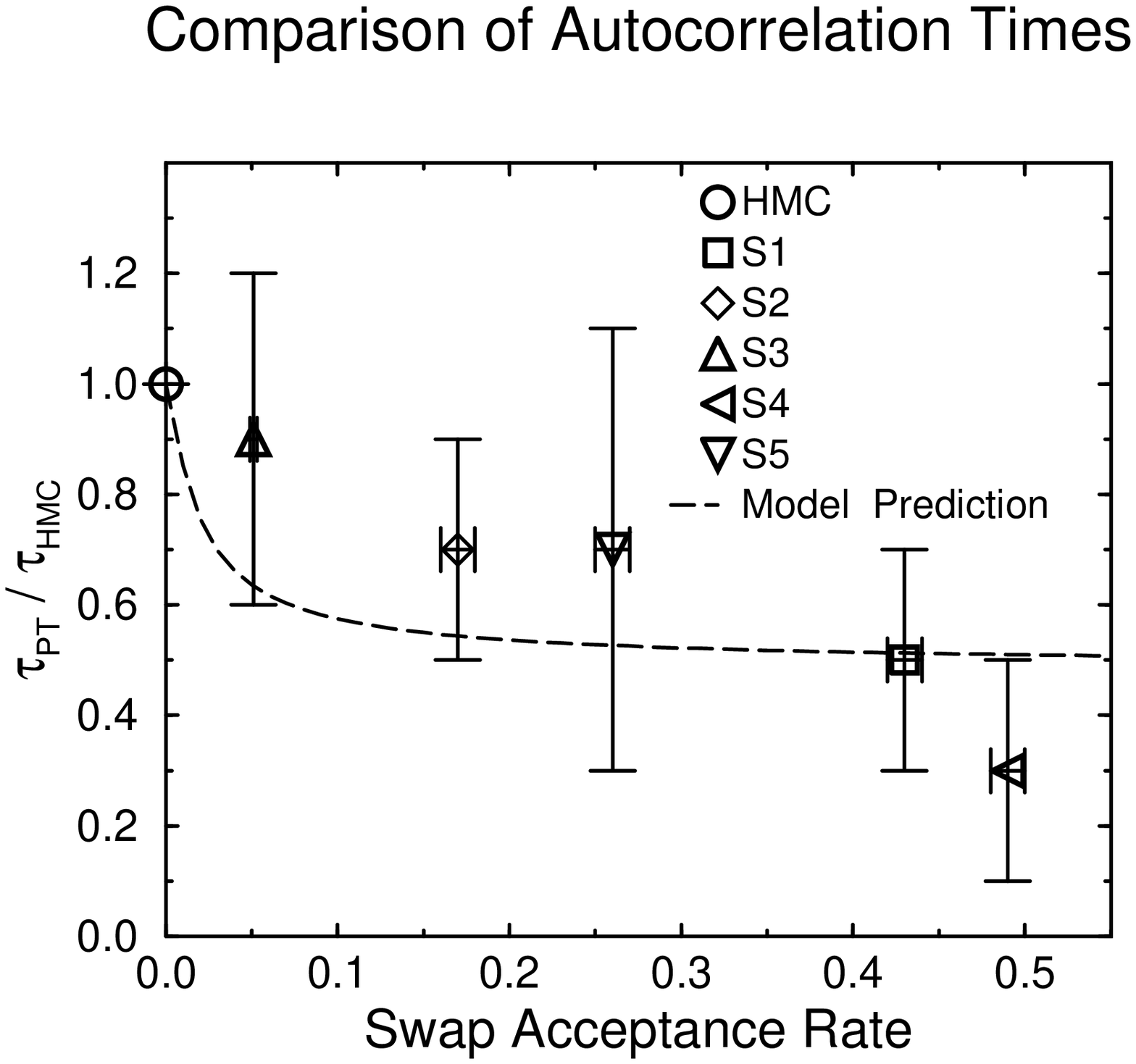}
\caption{Integrated autocorrelation times for the plaquette normalised
by that from GHMC simulations}
\label{f:norm_plaq_autocorr}
\end{figure}

Figure \ref{f:norm_plaq_autocorr} shows the ratio of PT to HMC autocorrelation times. The errors on the ratios were obtained by simple error combination. The line superimposed on the data in figure \ref{f:norm_plaq_autocorr} is the prediction of the model in section \ref{s:Algorithm} (c.f. equation
\ref{e:RATIO_PRED}). It can be seen that it is not inconsistent with the data.


\section{Summary and Conclusions}
\label{s:Conclusions}
In this paper we presented our study of the Parallel Tempering 
algorithm applied to lattice QCD with O(a)--improved Wilson fermions.
We showed how the algorithm satisfies detailed balance, and gave 
a formula for the swap acceptance rate in terms of the swap energy
change $\Delta \Ha$. We highlighted the connection of parallel tempering
with the technology of action matching. 
We presented and discussed a simple model of 
autocorrelations in a twin sub--ensemble PT system, and found that the 
algorithm is unlikely to improve autocorrelation times by more than a factor 
of two for such a system. We verified our simple model 
assumptions by gathering autocorrelation time data from previous simulations.

We carried out a numerical study where we verified the acceptance 
formula and the predictions of the autocorrelation model 
within statistical errors. We also obtained information on how 
the acceptance rate of the algorithm falls with increasing $\Delta \kappa$.

We found that using the pseudofermions from HMC on the swap attempt 
is a poor way to proceed if the parameters are matched for the $\mbox{Tr}\,\ln$ 
action. We have shown analytically that there is an extra noise term
in the definition of the distance between actions when pseudofermions are used.
We have attempted to estimate the size of this noise term numerically.

We conclude that Parallel
Tempering does not seem to give any real gain over HMC at the present
time.  We were unable to use PT to simulate sub--ensembles sufficiently
far apart in parameter space. The acceptance rate drops too quickly with
$\Delta \kappa$. This situation could be alleviated somewhat if the 
swap action/energy differences were to be calculated using the ${\rm Tr}\,\ln$ 
action, for simulations with parameters matched with that action. However
in the end the real problem is that the swap action/energy change scales
with the volume for a fixed kappa, and that when employing the PT algorithm 
on a realistic sized (eg $16^3 \times 32$) lattice, the scaling 
of the swap energy change would lower the acceptance rate and lose all 
that could be gained by using the $\mbox{Tr}\,\ln$ action.  

Thus we could not take advantage of the fact
that in one region of parameter space autocorrelation times are short
while in
 another they are long. With our parameter values, the HMC
autocorrelation times of our sub--ensembles are the same within
experimental errors and the predictions of our model apply. A chain of
sub--ensembles that would span the required distance in parameter space
can be constructed, but would take an unfeasibly large number of
sub--ensembles for lattices of interesting size.

\acknowledgements
We gratefully acknowledge support from PPARC grant no GR/L22744, and
EPSRC for funding under grant number GR/K41663.  We also wish to thank
Tony Kennedy and Stephen Booth for helpful discussions.

\newpage
\begin{table}
\caption{Simulation parameters used for twin ensemble runs and the reference HMC run}
\begin{tabular}{lcc}[h]
Simulation & ($\beta_1$, $c_{1}$, $\kappa_1$) & ($\beta_2$, $c_2$, $\kappa_2$) \\
\hline 
${\rm HMC}$     & $(5.2, 2.0171, 0.133)$ &    \\ 
\hline
$S1$     & $(5.2, 2.0171, 0.133)$ & $(5.2060, 2.01002, 0.13280)$ \\
$S2$     & $(5.2, 2.0171, 0.133)$ & $(5.2105, 2.00471, 0.13265)$ \\
$S3$     & $(5.2, 2.0171, 0.133)$ & $(5.2150, 1.99940, 0.13250)$ \\
\hline  
$S4$     & $(5.2, 2.0171, 0.133)$ & $(5.2, 2.0171, 0.13280)$ \\
$S5$     & $(5.2, 2.0171, 0.133)$ & $(5.2, 2.0171, 0.13265)$ \\
\end{tabular}
\label{t:2PARAMS}
\end{table}

\begin{table}[h]
\caption{Results from the PT simulations showing the appropriate results from HMC for comparison}
\begin{tabular}{lcccccc}
\mbox{Simulation} & $\Delta \beta (\times 10^{-3})$ & $\Delta \kappa (\times 10^{-4})$ & $\Expt{\Delta
\Ha}$ & $\Expt{A}$ & $\tau_1$ & $\tau_1/\tau_{\rm HMC}$ \\ \hline 
${\rm HMC}$  & &     &           &           & $26(6)$ & $1$ \\ \hline 
$S1$ & $6$    & $-2.0$ & $1.23(2)$ & $0.43(1)$ & $12(3)$ & $0.5(2)$ \\ 
$S2$ & $10.5$ & $-3.5$ & $3.76(4)$ & $0.17(1)$ & $19(4)$ & $0.7(2)$\\ 
$S3$ & $15$   & $-7.5$ & $7.64(6)$ & $0.051(2)$ & $24(6)$ & $0.9(3)$ \\ \hline 
$S4$ & $0$ & $-2.0$ & $0.91(4)$ & $0.49(1)$ & $9(4)$ & $0.3(2)$ \\ 
$S5$ & $0$ & $-3.5$ & $2.29(7)$ & $0.26(2)$ & $18(10)$ & $0.7(4)$ 
\end{tabular}
\label{t:MainRes}
\end{table}

\begin{table}
\caption{Comparison of $\mbox{Tr}\,\ln$ matching and acceptance with pseudofermionic acceptance}
\begin{tabular}{ccccc} 
Simulation & $\sigma^2(\Delta_{12})_{{\rm Tr}\,\ln}$ & $\sigma^2(\Delta)_{\rm p.f} = \Expt{\Delta \Ha}$ & $\langle A \rangle_{{\rm Tr}\,ln}$ & $\langle A \rangle_{\rm p.f}$ \\ \hline
$S3$ & $1.02(20)$ & $ 7.64(6)$ & $0.48(5)$ & $0.051(2)$ 
\end{tabular}
\label{t:Comppf}
\end{table}

\begin{table}[h]
\caption{The integrated autocorrelation times of some other simulations.}
\begin{tabular}{lccc}
 $\beta$ & $c$      & $\kappa$ & $\tau_{\rm HMC}$ \\ \hline
 $5.2$   & $1.99$   & $.1335$  & $18(8)$ \\ 
 $5.2$   & $2.0171$ & $.1330$  & $26(6)$ \\
 $5.232$ & $1.98$   & $.1335$  & $20(6)$ 
\end{tabular}
\label{t:IndepHMC}
\end{table}

\newpage

\begin{thebibliography}{99}
\frenchspacing
%
\bibitem{BJPHmc} S.~Duane, A.~D.~Kennedy, B.~J.~Pendleton, D.~Roweth, {\em Phys.\ Lett.}\ {\bf B195} (1987) 216-222.
%
\bibitem{Boyd1} B.~All\'es, G.~Boyd, M.~D'Elia, A.~Di~Giacomo, E.~Vicari, {\em Phys.Lett.} {\bf B389} (1996) 107-111, {\em hep-lat/9600749}
%
\bibitem{SESAM} B.~All\'es, G.~Bali, M.~D'Elia, A.~Di~Giacomo, N.~Eicker,
S.~G\"uesken, H.~H\"oeber, Th.~Lippert, K.~Schilling, A.~Spitz, T.~Struckmann, 
P.~Ueberholz, J.~Viehoff, {\em hep-lat/9803008}.
%
\bibitem{Hukushima} K.~Hukushima, J.~Nemoto, {\em cond-mat/9512035.}
%
\bibitem{Marinari} E.~Marinari, {\em cond-mat/9612010}
%
\bibitem{MariPari} E.~Marinari, G.~Parisi, J.~Ruiz-Lorenzo, {\em Spin Glasses and Random Fields}, edited by P. Young, {\em cond-mat/9701016}
%
\bibitem{Boyd2} G.~Boyd, {\em Nucl. Phys. (Proc. Suppl.)} {\bf 60A} (1998) 341-344, {\em hep-lat/9712012}
%
\bibitem{Clover} B. Sheikholeslami and R. Wohlert, 
        {\em Nucl.\ Phys.\ } {\bf B259} (1985) 572.
%
\bibitem{Jansen} K.~Jansen and R.~Sommer, {\em Nucl. Phys. B. (Proc. Suppl.)} {\bf63A-C} (1998) 853-855, {\em hep-lat/9709022}
%
\bibitem{ACIJCS} A.~C.~Irving and J.~C.~Sexton, {\em Phys. Rev.} {\bf D55} (1997) 5456.
%
\bibitem{ACIlat97} A.~C.~Irving, J.~C.~Sexton and E.~Cahill, {\em Nucl. Phys. B. (Proc. Suppl.)} {\bf 63A-C} (1998) 967
%
\bibitem{Match} A.~C.~Irving, J.~C.~Sexton, E.~Cahill, J.~Garden, 
        B.~Jo\'o, S.~M.~Pickles and Z.~Sroczynski, UKQCD Collaboration, 
        {\em hep-lat/9807015}
%
\bibitem{R0} R.~Sommer, Nucl.\ Phys.\ {\bf B411} (1994) 839.
%
\bibitem{Metropolis} N.~Metropolis, A.~W.~Rosenbluth, M.~N.~Rosenbluth, A.~H.~Teller, E.~Teller, {\em J. Chem. Phys} {\bf 21} (1953) 1087
%
\bibitem{Karsch} S.~Gupta, A.~Irb\"ack, F.~Karsch and
B. Petersson. {\em Phys. Lett.} {\bf B242} (1990) 437
%
\bibitem{Adk} I.~Horv\'ath, and A.~D.~Kennedy, {\em Nucl.\ Phys.\ } {\bf B510} (1998) 367-400, {\em hep-lat 9708024}
\bibitem{Sokal} N.~Madras, A.~D.Sokal, {\em J. Stat. Phys.} {\bf 50}
(1988) 109-186 
%
\bibitem{StephsAndZbysh} Z.~Sroczynski, S.~M.~Pickles and S.~P.~Booth, UKQCD Collaboration, {\em Nucl. Phys. B. (Proc. Suppl.)} {\bf 63A-C} (1998) 949-951.
\end{thebibliography}
\end{document}